\documentclass[showkeys,preprint,showpacs]{revtex4}
\begin{document}

\title{Brane Formation and Cosmological Constraint on the Number of Extra
Dimensions}

\author{Feng Luo\footnote{
fluo@student.dlut.edu.cn} and Hongya Liu\footnote{
hyliu@dlut.edu.cn}}

\address{Department of Physics, Dalian University of Technology,
Dalian, 116024, P. R. China}

\keywords{Extra dimensions; Special relativity; Brane.}
\pacs{11.10.Kk, 98.80.Cq, 11.25.-w, 11.27.+d}

\begin{abstract}
Special relativity is generalized to extra dimensions and quantized energy
levels of particles are obtained. By calculating the probability of
particles' motion in extra dimensions at high temperature of the early
universe, it is proposed that the branes may have not existed since the very
beginning of the universe, but formed later. Meanwhile, before the
formation, particles of the universe may have filled in the whole bulk, not
just on the branes. This scenario differs from that in the standard big bang
cosmology in which all particles are assumed to be in the 4D spacetime. So,
in brane models, whether our universe began from a 4D big bang singularity
is questionable. A cosmological constraint on the number of extra dimensions
is also given which favors $N\geq 7$.
\end{abstract}

\maketitle

\section{Introduction}

In the Kaluza-Klein (K-K) theory \cite{Overduin}, and the later developed
string theory \cite{Davies}, \cite{Schwarz}, the scale of the extra
dimensions is the Planck scale ($\sim 10^{-35}m$), so the reason why we can
not see the extra dimensions, or why the particles can not run into the
extra dimensions, is simply that the particles' de Broglie wave lengths are
much larger than the scale of the extra dimensions even in the most powerful
accelerators. Almost everyone is satisfied with this explanation and no
further researches about the particles' motion in the extra dimensions are
needed to be done, since the huge particles can not even \textquotedblleft
smell\textquotedblright\ the tiny dimensions. That is, the energy required
to detect Planck scale---the Planck energy ($\sim 10^{19}GeV$)---can never
be reached in the accelerators in the expected future. However, from the end
of 1990s, the possibility of large extra dimensions (even up to the
sub-millimeter) has been proposed in some \textquotedblleft brane
world\textquotedblright\ theories (e.g. \cite{Arkani-Hamed}, \cite%
{Antoniadis}, \cite{Arkani}), and the possible phenomena in the coming
accelerators (like the NLC and LHC) relate to the extra dimensions have been
predicted. Many experimentalists are eager to receive the first signal
coming from extra dimensions.

Of course, since the standard model (SM) has passed numerous tests without
deviation, a feasible \textquotedblleft brane world\textquotedblright\ model
with large extra dimensions must consider the method of localizing the SM
particles on the branes, and indeed many methods have been given (e.g. \cite%
{Dvali}, \cite{Neronov}, \cite{Dubovsky}, \cite{Dva}).

Since the scale of extra dimensions may be much larger than the Planck
scale, if we do not want to be bothered to struggle with those complicated
localizing methods, but only consider the simple restraint from de Broglie
wave lengths, there will really be some possibilities for the particles to
jump into the relatively larger extra dimensions and to move in them. We
would like to pointed out that the possibility of the particles entering
extra dimensions has already been discussed before, like in the influential
paper presented by N.Arkani-Hamed, S.Dimopoulos and G.Dvali (ADD) \cite%
{Arkani-Hamed}. In their framework, the SM particles can be kicked into
extra dimensions with sufficiently hard collisions, carrying away energy,
orbiting around the extra dimensions. And if the topology of the extra
dimensional compact manifold is appropriate, the particles will periodically
return to our four-dimensional spacetime. However, all of these kinds of
descriptions are just discussed in rough forms, the fundamental physical
framework of the particles' motion in extra dimensions has not been studied
systematically, especially in a mathematical form. And we all know that
after setting up the basic frame (although may seem simple and clear), some
physical meanings of this kind of motion can be considered more in detail.
Most important of all, a clear framework may provide us some new standpoints
about extra dimensions. So it is deserved to research this possible kind of
motion seriously.

In this paper, we will generalize the special relativity to extra dimensions
and get the quantized energy levels of the particles. According to the
energy levels, the Boltzmann's distribution of the particles' number density
will be given. We will analysis that this framework leads to the problem of
brane formation, and consequently, a constraint for the number of extra
dimensions is proposed from the consideration of cosmology.

\section{Extra dimensions and the quantized energy levels}

The initial theory about extra dimensions (K-K theory) just assumed that
they are very small and compacted. However, some new properties were added
to extra dimensions in the subsequent theories which naturally became more
complicated. We would like to recover the original simple assumption and
believe the extra dimensions are really physically real, that is they are
not just visual aids introduced to describe the theories. Under this
assumption, the extra dimensions may be equally treated as the ordinary
three spaces, except that their topologies and scales are quite different.
Naturally, the theories of four-dimensional spacetime can be generalized to
and may be still suitable to describe the $(N+4)$ dimensional spacetime.

Suppose a particle with a non-zero rest mass can move in the extra
dimensions, then the four-dimensional special relativity can be generalized
to $(N+4)$ dimensions, where $N$\ is the number of extra dimensions. The
familiar formula about energy is generalized as
\begin{equation}
E^{2}=p^{2}c^{2}+m_{0}^{2}c^{4}=\mathbf{p}_{3}^{2}c^{2}+%
\sum_{i=1}^{N}p_{Ei}^{2}c^{2}+m_{0}^{2}c^{4},  \label{eq1}
\end{equation}%
where $\mathbf{p}_{3}\ $is the momentum of the particle in the ordinary
three spaces, $p_{Ei}$\ is its momentum in the $i$th extra dimension. We
have already supposed that the extra dimension, between them and the
ordinary three spaces, are orthogonal. Like the general assumption in the
string theories, we also assume that the extra dimensions are compact.
Furthermore, in order to make the discussion simple, we suppose the $N$\
extra dimensions are all circular shape.

From the familiar de Broglie relation, we write the momentum $p_{Ei}$ as
\begin{equation}
p_{Ei}=\frac{h}{\lambda _{Ei}}, \quad i=1,2,\cdots ,N,
\label{eq2}
\end{equation}%
where $\lambda _{Ei}$ is the projection of de Broglie wave length in the $i$%
th extra dimension, $h$ is the Planck constant. The momentum $\mathbf{p}%
_{3}\ $still satisfies

\begin{equation}
\mathbf{p}_{3}=\frac{E\mathbf{v}_{3}}{c^{2}},  \label{eq3}
\end{equation}%
where $\mathbf{v}_{3\text{ }}$is the speed in the ordinary three spaces.

Because we suppose the $i$th extra dimension is a ring, we can use the
stationary wave requirement
\begin{equation}
\lambda _{Ei_{j}}=\frac{2\pi r_{i}}{\left\vert j_{i}\right\vert }=\frac{b_{i}%
}{\left\vert j_{i}\right\vert },\quad j_{i}=\pm 1,\pm 2,\cdots ,  \label{eq4}
\end{equation}%
where $r_{i}$ is the radius of $i$th extra dimension and $b_{i}$ is its
perimeter. The signs of $j_{i}$ indicate that the particle can both circle
clockwise and anticlockwise directions on the rings. If $j_{i}=0$, $\lambda
_{Ei_{0}}$ will be $\infty $, which means the particle can not see the $i$th
extra dimension and it has no motion in it. Put Eq. (\ref{eq2}), (\ref{eq3}%
), (\ref{eq4}) to Eq. (\ref{eq1}), we get
\begin{equation}
E_{j_{1},j_{2},\cdots ,j_{N}}=\frac{m_{0}c^{2}\sqrt{1+\sum_{i=1}^{N}\frac{%
j_{i}^{2}h^{2}}{m_{0}^{2}b_{i}^{2}c^{2}}}}{\sqrt{1-\frac{v_{3}^{2}}{c^{2}}}}.
\label{eq5}
\end{equation}%
Since the momentum in the $i$th extra dimension can be written as%
\begin{equation}
p_{Ei}=\frac{m_{0}v_{Ei}}{\sqrt{1-\frac{v_{3}^{2}+\sum\limits_{k=1}^{N}{v}%
_{Ek}^{2}}{c^{2}}}},  \label{eq6}
\end{equation}%
where $v_{Ei}$ is the particle's speed in the $i$th extra dimension, then
from Eq. (\ref{eq2}), (\ref{eq4}) and (\ref{eq6}), we get
\begin{equation}
v_{Ei_{j}}=\frac{\left\vert j_{i}\right\vert h\sqrt{1-\frac{v_{3}^{2}}{c^{2}}%
}}{m_{0}b_{i}\sqrt{1+\sum_{k=1}^{N}\frac{j_{k}^{2}h^{2}}{%
m_{0}^{2}b_{k}^{2}c^{2}}}}.  \label{eq7}
\end{equation}%
We can also obtain the period of the particle returning to the
four-dimensional spacetime%
\begin{equation}
T_{Ei_{j}}=\frac{b_{i}}{v_{Ei_{j}}}.  \label{eq8}
\end{equation}%
Clearly, because of the motion in the extra dimensions, the particle's
energy is quantized. Now we introduce the concept of energy levels. We call $%
E_{0}$ the ground state energy, $E_{\pm 1}$, $E_{\pm 2}\cdots $ the first,
second \ldots\ excited states energies. Obviously, the excited energies are
at least double degenerated under the assumption of this kind of topological
structure. \textquotedblleft At least\textquotedblright\ here means if all
the extra dimensions really share the same radius, as usually expected, the
degenerations of excited energy levels will be much larger. For example, the
first excited energy level contains $2^{1}C_{N}^{1}$ states, the fourth
excited energy level contains ($2^{4}C_{N}^{4}+2^{1}C_{N}^{1}$) states.

We should point that for the particles, to be in excited states and to run
into extra dimensions is the same thing in our framework. In other words,
the motion in extra dimensions can be described by the labels of excited
states.

Evidently, when $j_{i}=0$ ($i=1,2,\cdots ,N$), the ground state energy is
\begin{equation}
E_{0}=\frac{m_{0}c^{2}}{\sqrt{1-\frac{v_{3}^{2}}{c^{2}}}},  \label{eq9}
\end{equation}%
and $v_{Ei_{0}}=0$, ($i=1,2,\cdots ,N$), which corresponds to the usual
four-dimensional special relativity.

The concept of quantized energy (or say, the quantized mass, since $E=mc^{2}$%
) is not strange to those who are familiar with the Kaluza-Klein (K-K)
theory \cite{Overduin}. The above deduction seems quite simple and the
generalization of special relativity looks as if nearly nothing more than
standard. However, when considering other requirements, like the thermal
statistical theory, this framework can really lead to some interesting
results.

The statistical theory could also be generalized to extra dimensions since
we have assumed that the extra dimensions share the same qualities with the
ordinary three spaces except the topology and scales. Then the distribution
of the particles' number density in different energy levels should obey the
Boltzmann's distribution law (for simplicity, we will not consider the
deviation from Boltzmann's distribution because of the particles' identical
property, and we assume the system of the particles are in the equilibrium
state)
\begin{equation}
\frac{\rho _{E_{j_{1},j_{2},\cdots ,j_{N}}}}{\rho _{E_{0}}}=\varpi
_{E_{j_{1},j_{2},\cdots ,j_{N}}}\exp \left( {\ -\frac{\Delta
E_{j_{1},j_{2},\cdots ,j_{N}}}{kT}}\right) ,  \label{eq10}
\end{equation}%
where $k$ is the Boltzmann's constant, $T$ is the temperature of the system,
$\varpi _{j_{1},j_{2},\cdots ,j_{N}}$ is the degeneration, and $\Delta
E_{j_{1},j_{2},\cdots ,j_{N}}=E_{j_{1},j_{2},\cdots ,j_{N}}-E_{0}$ is the
difference between the energy of the excited and the ground levels.

In the following discussion, we use the units $c=\hbar =k=1$. Then Eqs. (\ref%
{eq5}), (\ref{eq7}) and (\ref{eq10}) take the form
\begin{equation}
E_{j_{1},j_{2},\cdots ,j_{N}}=\frac{m_{0}\sqrt{1+\sum_{i=1}^{N}\frac{%
j_{i}^{2}}{m_{0}^{2}r_{i}^{2}}}}{\sqrt{1-v_{3}^{2}}},  \label{eq11}
\end{equation}%
\begin{equation}
v_{Ei_{j}}=\frac{\left\vert j_{i}\right\vert \sqrt{1-v_{3}^{2}}}{m_{0}r_{i}%
\sqrt{1+\sum_{k=1}^{N}\frac{j_{k}^{2}}{m_{0}^{2}r_{k}^{2}}}},  \label{eq12}
\end{equation}%
\begin{equation}
\frac{\rho _{E_{j_{1},j_{2},\cdots ,j_{N}}}}{\rho _{E_{0}}}=\varpi
_{j_{1},j_{2},\cdots ,j_{N}}\exp \left( {\ -\frac{\Delta
E_{j_{1},j_{2},\cdots ,j_{N}}}{T}}\right) .  \label{eq13}
\end{equation}%
Since for the particles, to be in excited states and run into the extra
dimensions is the same thing, Eq. (\ref{eq13}) is actually the formula to
estimate the probability for the particles to enter extra dimensions.

\section{Brane formation and the constraint on the number of extra dimensions%
}

From Eq. (\ref{eq13}), one can notice that when the temperature $T$ is high
enough, the particles may have motion in the extra dimensions. Considering
the early universe was in the thermodynamic equilibrium state with extremely
high temperature, thus a great part of particles may really have rushed in
the extra dimensions, and then, with the decrease of temperature, they fell
down to the ground state, that is, down to the ordinary three spaces.

In other words, the brane on which we live may have not existed since the
very beginning of the universe, but formed later. And before the formation
of the branes, particles of the universe may distributed in the whole $(N+4)$
dimensional bulk, not just on the branes. This kind of scenario is clearly
different from that in the standard 4D cosmological model in which it is
assumed that all particles of the universe are located in the 4D spacetime.
So, in brane models, whether our universe began from a big bang or other
kind of singularities, such as in the ekpyrotic/cyclic \cite{Steinhardt} or
in the big crunch/ big bounce models \cite{Claus} \cite{Wesson}, becomes
unclear and deserves more serious studies.

The definite description of the above suggestion needs the definite
information about the topology and scales of extra dimensions. For the scale
of extra dimensions, we would like to take the formula proposed by ADD \cite%
{Arkani-Hamed},

\begin{equation}
R_{N}\sim 10^{\frac{^{30}}{N}-17}cm\times (\frac{1TeV}{m_{EW}}),
\label{eq14}
\end{equation}%
where $m_{EW}$ is the electroweak scale $m_{EW}\sim 1TeV$. A more convenient
formula can be written as

\begin{equation}
R_{N}\sim 10^{\frac{^{30}}{N}-(3\sim 4)}GeV^{-1}.  \label{eq15}
\end{equation}%
Since the topology of extra dimensions assumed by ADD is the same to ours,
we can safely use this formula to make a discussion.

First, we define the time $t_{1}$ when $T\sim \Delta E_{1}$ (corresponding
to $\frac{\rho _{E_{1}}}{\rho _{E_{0}}}\sim \exp (-1)$) and the time $t_{2}$
when $T\sim 10^{-1}\Delta E_{1}$ (corresponding to $\frac{\rho _{E_{1}}}{%
\rho _{E_{0}}}\sim \exp (-10)$) respectively as the start and end moment of
the formation of branes. We aware that $t_{2}$ should not be later than $1$
second or so, since the knowledge about the evolution of the early universe
after that time is quite credible (e.g. The standard FRW model says that the
decay of free neutrons started at about $1s$ and the Primordial
Nucleosynthesis started at about $10^{2}s$, the predictions of which agree
with the observation very well). However, the evolution of the universe in
the whole ($N+4$) spacetime may be different with the one on the branes. So
the extra dimensions should not play any important role at least on the
early evolution after $1$ second. In the following discussion, we assume
that the topology and scale of the extra dimensions are fixed or at least
had already fixed.

The quark-hadron transition took place when $t\sim 10^{-4}s$, which
corresponds to $T\sim 10^{-1}GeV$ (Since from the standard FRW model, $%
T(MeV)\sim t^{-\frac{1}{2}}(s)$. But one should notice that the use of this
relation is just to make description easy. Actually, the description of the
evolution with temperature is more appropriate, since we aware that this
simple relation may not hold when consider the evolution of the extremely
early universe with extra dimensions). The rest mass of a nucleon is $%
m_{0}\approx 1GeV$, thus the motion of these particles when they were born
was non-relativistic $v_{3}\ll 1$. Using the relations mentioned above, we
can easily obtain the relationship between $t_{1}$ and $R_{N}$ as
\begin{equation}
t_{1}\sim 10^{-6}\left[ m_{0}(\sqrt{1+\frac{1}{m_{0}^{2}R_{N}^{2}}}-1)\right]
^{-2},  \label{eq16}
\end{equation}%
and also we have
\begin{equation}
t_{2}\sim 10^{-4}\left[ m_{0}(\sqrt{1+\frac{1}{m_{0}^{2}R_{N}^{2}}}-1)\right]
^{-2}.  \label{eq17}
\end{equation}%
Notice that these expressions are suitable to any non-relativistic
particles. Then

for $N=1$, $t_{1}\sim 10^{101}s$;

for $N=4$, $t_{1}\sim 10^{11}s$;

for $N=5$, $t_{1}\sim 10^{5}s$.

Even for $N=6$, $t_{1}\sim 10^{1}s$, $t_{2}\sim 10^{3}s$, which is still too
late. Luckily, for $N=7$, which is also the favorite number in the M/string
theories, the corresponding $t_{1}\sim 10^{-2}s$ and $t_{2}\sim 1s$, which
is just permissible for the later evolution of the universe.

Notice that for $N=7$ and $T\sim 10^{-1}GeV$, we have $\Delta E\sim
10^{-2}GeV$ and $\frac{\rho _{E_{1}}}{\rho _{E_{0}}}\sim \exp
(-10^{-1})\approx 1$, and the same order is also suitable for not too high
excited energy levels. Moreover, when considering the degenerations, we can
come to a conclusion that the newly born nucleons actually scudded in the
whole $(N+3)$ spaces. That is, the branes had not formed by then. The branes
began to form when $T\sim \Delta E_{1},$ the feasibility for this definition
can also be shown in the following way.

Since $T\ll m_{0}$, the formula of the particles' average speed from
Boltzmann' statistics can be used%
\begin{equation}
v_{3}=\sqrt{\frac{8kT}{\pi m_{0}}}\sim \sqrt{\frac{T}{m_{0}}}\text{ \ (}k=1%
\text{)}.  \label{eq18}
\end{equation}%
Notice that for $N\leq 7$, $\frac{1}{m_{0}^{2}R_{N}^{2}}\leq 1$ is
satisfied, so $\Delta E_{1}\approx \frac{1}{2m_{0}R_{N}^{2}}$ and

\begin{equation}
v_{E_{1}}=\frac{\sqrt{1-v_{3}^{2}}}{m_{0}R_{N}\sqrt{1+\frac{1}{%
m_{0}^{2}R_{N}^{2}}}}\approx \frac{1}{m_{0}R_{N}}.  \label{eq19}
\end{equation}%
When put $T\sim \Delta E_{1}$ to Eq. (\ref{eq18}), we can find $v_{3}\sim
v_{E_{1}}$, that is the random velocity in three spaces share the same order
with the first few quantized speeds in extra dimensions. Furthermore, when $%
T>\Delta E_{1}$, $v_{3}>v_{E_{1}}$, which indicates that the particles
actually easily run in the extra dimensions. When $T\ll \Delta E_{1}$, $%
v_{3}\ll v_{E_{1}}$, then it is hard for the particles to step into the
extra dimensions.

Now let's summarize the description above. Nucleons were born when $t\sim
10^{-4}s$, and they can see the whole $(N+3)$ spaces at that time. At $t\sim
10^{-2}s$, they began to jump down to the ground state and the branes began
to form. With the decrease of the temperature, more and more nucleons fell
down to the ground state. Finally, at $t\sim 1s$, the brane we live on
formed completely, and the extra dimensions stepped down from the stage,
then the familiar evolution of the universe began.

We should point that the description is on the assumption of $N=7$, or say,
the scale of the extra dimensions $R_{N}\sim 10^{-15}m$. If $N>7$, the scale
is shorter, the corresponding $\Delta E_{1}$ is larger, and $t_{1}$ is
earlier. If we also take $m_{0}\sim 1GeV$, then for $N=8$, $\Delta E_{1}\sim
10^{-2}GeV$, $t_{1}\sim 10^{-4}s$; for $N=9$, $\Delta E_{1}\sim 10^{-1}GeV$,
$t_{1}\sim 10^{-6}s$. Notice that when $N\geq 9$, $t_{1}$ is earlier than
the moment of quark-hadron transition, so the use of $m_{0}\sim 1GeV$ is not
appropriate. However, if we believe that the branes formed after the birth
of the nucleons, only $N=7$ and $N=8$ are proper. To go to a step further,
if the parameters, like $m_{EW}$ and the time of quark-hadron transition are
more exactly fixed, maybe only one choice is permissible.

\bigskip

Of course, the possibility that the branes had formed before the time of
quark-hadron transition can not be excluded curtly. However, a definite
discussion requires the knowledge of the very early universe, that is the
knowledge before $10^{-4}s$. Since we have not possessed the exact
information about that time by now, we could only give a rough description.
But if the particles at that time were all relativistic, the following
description is really credible.

For relativistic particles, $T\sim E$, since $E^{2}\approx \mathbf{p}%
_{3}^{2}+\sum_{i=1}^{N}p_{Ei_{j}}^{2}$, if we require $p_{3}\sim
p_{E_{N_{1}}}$ as the mark of the beginning formation of branes, the
rationality of which can be easily understood from the consideration of the
physical meaning, then we have $T_{crit}\sim p_{E_{N_{1}}}$, where $T_{crit}$
is the temperature when the branes began to form. When using Eq. (\ref{eq2})
and (\ref{eq4}), we can obtain the relationship between $T_{crit}$ to the
scale of extra dimensions $R_{N}$%
\begin{equation}
T_{crit}\sim \frac{1}{R_{N}}.  \label{eq20}
\end{equation}%
So from Eq. (\ref{eq15}), we have

for $N=1$, $\ T_{crit}\sim 10^{-27}GeV$;

for $N=5$, $\ T_{crit}\sim 10^{-3}GeV$;

for $N=6$, $\ T_{crit}\sim 10^{-2}GeV$;

for $N=7$, $\ T_{crit}\sim 10^{-1}GeV$;

for $N=10$, $T_{crit}\sim 10^{0}GeV$.

Notice that $T_{crit}$ must greater than $10^{-3}GeV$, which corresponds to $%
1s$. Furthermore, considering for $N\leq 6$, the corresponding $T_{crit}$
are lower than the generation temperature of nucleons, actually they have
already been excluded by the discussion above. So only $N\geq 7$ satisfies
our requirement. However, no upper limit can be provided this time, since no
other information (like the definite type of particles) is available.

We should point out that whatever the detail is, the tenor we proposed is
that the branes may have not existed since the very beginning of the
universe, but formed later. Because the particles gradually lost the ability
of running in the extra dimensions with the decrease of temperature, they
jumped down to the ground state but can not jump up again, thus the branes
formed. Moreover, the critical moment or the critical temperature of the
formation of branes depends on the scale of extra dimensions, thus here
comes a new constraint for the number of extra dimensions. Beside these, the
extra dimensions may have played an important role on the evolution of the
extremely early universe before the branes had formed. So a serious study of
the extra dimensions may shed light on the research of the infant universe.

\section{Conclusion}

We generalized the four-dimensional special relativity to the
$(N+4)$ dimensions. In the circumstance of circular-shape extra
dimensions, we gave the quantized energies of the particles. We
introduced the concept of energy levels, and linked the motion in
extra dimensions to the excited energy states. In this framework,
and based on the formula about the scale of extra dimensions given
by ADD, the problem of brane formation was proposed. That is the
branes may has not existed from the very beginning of the
universe, but was generated later. Consequently, before the
formation of the branes, particles of the universe may fill in the
whole bulk, the evolution of the universe may differ from that in
the 4D standard model, and even the nature of the big bang
singularity could be changed. A new constraint for the
number of extra dimensions was also derived, and the result favored $N\geq 7$%
. We believe that the extra dimensions may have played a significant role on
the evolution of the extremely early universe and this deserves more further
studies.

\section*{Acknowledgments}

{This work was supported by NSF (10273004) and NBRP (2003CB716300) of P. R.
China.}

\end{document}